# Carrier Phase Estimation in Dispersion-Unmanaged Optical Transmission Systems

Tianhua Xu, Polina Bayvel
Dept. Electronic & Electrical Engineering
University College London (UCL)
London, UK
e-mail: tianhua.xu@ucl.ac.uk

Tiegen Liu, Yimo Zhang
Dept. Optical Engineering
Tianjin University
Tianjin, China
e-mail: tgliu@tju.edu.cn

Gunnar Jacobsen, Jie Li
Networking and Transmission Lab
Acreo Swedish ICT AB
Kista, Sweden
e-mail: gunnar.jacobsen@acreo.se

Sergei Popov
Optics & Photonics Group
Royal Institute of Technology (KTH)
Stockholm, Sweden
e-mail: sergeip@kth.se

(Invited)

*Abstract*—The study on carrier phase estimation (CPE) approaches, involving a one-tap normalized least-mean-square (NLMS) algorithm, a block-wise average algorithm, and a Viterbi-Viterbi algorithm has been carried out in the long-haul high-capacity dispersion-unmanaged coherent optical systems. The close-form expressions and analytical predictions for bit-error-rate behaviors in these CPE methods have been analyzed by considering both the laser phase noise and the equalization enhanced phase noise. It is found that the Viterbi-Viterbi algorithm outperforms the one-tap NLMS and the block-wise average algorithms for a small phase noise variance (or effective phase noise variance), while the three CPE methods converge to a similar performance for a large phase noise variance (or effective phase noise variance). In addition, the differences between the three CPE approaches become smaller for higher-level modulation formats.

## I. Introduction

Optical fibre networks constitute the substantial part of current communication infrastructure by carrying most of digital traffic data. Long-haul high-capacity optical fibre networks can be seriously deteriorated by physical impairments in the transmission systems, e.g. chromatic dispersion (CD), polarization mode dispersion, laser phase noise (PN) and fibre nonlinearities [1-20]. Since the compensation and mitigation of these system impairments can be implemented in electrical domain, the combination of coherent optical detection, advanced modulation formats and digital signal processing (DSP) has become one of the most popular and promising solution for new-generation long-haul and metro optical communication networks to perform close to the limit of Shannon capacity, based on the knowledge of signal amplitude and phase [21-29]. To compensate the PN coming from the transmitter (Tx) and local oscillator (LO) lasers, some research has been carried out on the feed-forward and feed-back carrier phase estimation (CPE) to estimate the phase of optical carriers [30-38]. Among these, the one-tap normalized least-mean-square (OT-NLMS), the block-wise average (BWA), and the Viterbi-Viterbi (VV) CPE approaches have been justified for compensating the carrier PN effectively, which are considered as promising signal processing methods for next-generation high-capacity fibre communication systems [36-38].

In fibre communication networks employing DSP based CD compensation, a phenomenon so-called "equalization enhanced phase noise (EEPN)" is induced from the interplay between dispersion compensation module and carrier PN [39,40]. The optical fibre communication systems will be more severely degraded by EEPN, with the increase of chromatic dispersion, symbol rate, system bandwidth, laser 3-dB linewidth (Tx or LO), and modulation format [41-44]. The effects of EEPN have been studied for single-channel, WDM, OFDM, pre-distortion of dispersion, and multi-mode fibre transmission networks [45-49]. Also, research has been implemented to study the performance of EEPN in the CPE in long-haul optical transmission systems [50-56]. Considering the impact of EEPN, traditional analyses of the CPE algorithms are not appropriate any more for the design of DSP based high-capacity communication systems. Correspondingly, it would be very interesting and useful to study the BER performance in the OT-NLMS, the BWA, and the VV CPE approaches, when the impact of both laser PN and EEPN is considered.

In this paper, theoretical assessments on the CPE in long-haul fibre communication networks using the OT-NLMS, the BWA, and the VV CPE are presented and discussed. The close-form expressions of the estimated carrier PN within these CPE approaches are derived, and the BER performance, e.g. the BER floors, has been analytically investigated. For different PN variance (or effective PN variance considering

EEPN), the theoretical performance of the OT-NLMS, BWA, and VV carrier phase estimation has been compared. It is shown that the VV CPE method outperforms the OT-NLMS and the BWA CPE methods for a small PN variance (or effective PN variance), whereas three carrier phase estimation algorithms converge to a similar behavior for a large PN variance (or effective PN variance). Besides, the performance of the three CPE approaches gets closer with the increase of the level of modulation format.

## II. PRINCIPLE OF LASER PN AND EEPN

### A. Laser PN

The PN in Tx and LO lasers in optical fibre networks follow a Lorentzian distribution and the laser PN variance can be described as [1,2],

$$\sigma^2_{Tx\_LO} = 2\pi(\Delta f_{Tx} + \Delta f_{LO}) \cdot T_S \quad (1)$$

where $\Delta f_{Tx}$ and $\Delta f_{LO}$ are the 3-dB linewidths of Tx and LO laser respectively, and $T_S$ is the symbol period. Eq. (1) shows that the laser PN variance will decrease with the increase of symbol rate.

### B. EEPN

The PN variance of EEPN in long-haul optical networks follows the analytical description in Ref [39,44], where the interaction between the dispersion compensation and the LO laser PN is considered,

$$\sigma^2_{EEPN} = \Delta f_{LO} \cdot D \cdot L \cdot \pi \lambda^2 / 2cT_S \quad (2)$$

where $D$ is CD coefficient, $L$ is fibre length, $\lambda$ is wavelength of optical carrier wave.

When EEPN is considered in carrier phase estimation, total effective PN variance in the $n$-level PSK optical fibre networks can be expressed as the following description [41,44],

$$\begin{aligned}\sigma^2_T &\approx \sigma^2_{EEPN} + \sigma^2_{Tx\_LO} + 2\rho \cdot \sigma_{EEPN} \cdot \sigma_{Tx\_LO} \\ &\approx \sigma^2_{EEPN} + \sigma^2_{Tx\_LO} \\ &= \pi \lambda^2 D \cdot L \cdot \Delta f_{LO}/2cT_S + 2\pi T_S(\Delta f_{Tx} + \Delta f_{LO})\end{aligned} \quad (3)$$

## III. ANALYSIS OF CARRIER PHASE ESTIMATION

### A. OT-NLMS CPE

The transfer function of the OT-NLMS CPE in optical networks can be expressed as follows [35,44],

$$y(k) = w_{NLMS}(k)x(k) \quad (4)$$

$$w_{NLMS}(k+1) = w_{NLMS}(k) + \mu e(k)x^*(k)/|x(k)|^2 \quad (5)$$

$$e(k) = d(k) - y(k) \quad (6)$$

where $x(k)$ means input signal, $k$ means symbol index, $y(k)$ means output signal, $w_{NLMS}(k)$ represents $k$-th tap weight, $d(k)$ is desired output signal, $e(k)$ means the error between output signal and desired signal, and $\mu$ means the step size in OT-NLMS approach.

The OT-NLMS CPE method behaves similar to the ideal differential CPE, and the BER floor in the OT-NLMS CPE for QPSK optical networks can be approximated as [35,44]:

$$BER^{NLMS\_QPSK}_{floor} \approx \frac{1}{2} erfc\left(\frac{\pi}{4\sqrt{2}\sigma_T}\right) \quad (7)$$

Accordingly, the BER floor in OT-NLMS approach for $n$-level PSK optical fibre networks is calculated as follows:

$$BER^{NLMS}_{floor} \approx \frac{1}{\log_2 n} erfc\left(\frac{\pi}{n\sqrt{2}\sigma_T}\right) \quad (8)$$

where $\sigma^2_T$ is the total PN variance for $n$-level PSK networks.

### B. BWA CPE

For $n$-level PSK networks, the estimated carrier phase in each data block using the BWA approach is described as in Ref [35,37],

$$\phi_{BWA}(k) = \frac{1}{n} \arg\left\{\sum_{p=1+(q-1)\cdot N_{BWA}}^{q \cdot N_{BWA}} x^n(p)\right\} \quad (9)$$

$$q = \left\lceil \frac{k}{N_{BWA}} \right\rceil \quad (10)$$

where $k$ is signal index, $N_{BWA}$ is block size of BWA method.

The BER floor for BWA CPE in $n$-level PSK transmission networks is described using Taylor expansion as follows:

$$BER^{BWA}_{floor} \approx \frac{1}{N_{BWA} \log_2 n} \cdot \sum_{p=1}^{N_{BWA}} erfc\left(\frac{\pi}{n\sqrt{2}\sigma_{BWA}(p)}\right) \quad (11)$$

$$\sigma^2_{BWA}(p) = \frac{\sigma^2_T}{6N^2_{BWA}}\left[\begin{array}{l}2(p-1)^3 + 3(p-1)^2 + 2(N_{BWA}-p)^3 \\ + 3(N_{BWA}-p)^2 + N_{BWA} - 1\end{array}\right] \quad (12)$$

where $\sigma^2_T$ is the total PN variance in $n$-level PSK networks.

### C. VV CPE

The estimated carrier phase in VV method for the $n$-level PSK optical networks is described using the equation in Ref [35,38],

$$\phi_{VV}(k) = \frac{1}{n}\arg\left\{\sum_{q=-(N_{VV}-1)/2}^{(N_{VV}-1)/2} x^n(k+q)\right\} \quad (13)$$

where $N_{VV}$ is the block size of VV CPE algorithm.

The BER floor in the VV carrier phase estimation for $n$-level PSK coherent transmission networks is assessed using following description:

$$BER_{floor}^{VV} \approx \frac{1}{\log_2 n} erfc\left(\frac{\pi}{n\sigma_T}\sqrt{\frac{6N_{VV}}{N_{VV}^2-1}}\right) \quad (14)$$

where $\sigma_T^2$ is the total PN variance in $n$-level PSK networks.

## IV. RESULTS AND DISCUSSIONS

As shown from Fig. 1 to Fig. 5, the BER floors versus PN variances (or effective PN variance considering EEPN) in the above three (the OT-NLMS, the BWA, and the VV) CPE methods in long-haul fibre networks have been comparatively investigated, where the modulation formats of the QPSK (Fig. 1), the 8-PSK (Fig. 2), the 16-PSK (Fig. 3), the 32-PSK (Fig. 4) and the 64-PSK (Fig. 5) are applied respectively. In all these analytical models, the attenuation, the PMD, the fibre nonlinearities are neglected. A block size of 15 is used in both BWA and VV carrier phase estimation methods, since the AWGN in the channel such as amplified spontaneous emission (ASE) noise from the optical amplifiers should also be taken into consideration in practical optical communication systems. The BER floors in the three CPE approaches are evaluated and discussed comparatively in the range from $10^{-6}$ to 0.5.

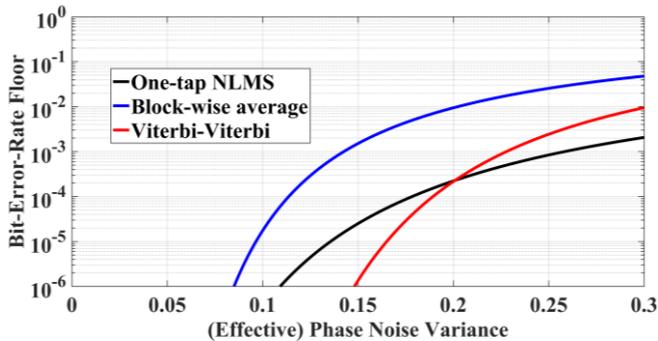

Fig. 1. BER floors versus PN variances in the three CPE approaches in QPSK optical fibre transmission network.

It can be found in Fig. 1 that, in the QPSK transmission system, the Viterbi-Viterbi CPE method outperforms the OT-NLMS and the BWA methods for a small PN variance (or effective PN variance), whereas the three CPE algorithms will converge to a similar behavior for a large PN variance (or effective PN variance). The same trends can be found in the 8-PSK optical network in Fig. 2, the 16-PSK optical network in Fig. 3, the 32-PSK optical network in Fig. 4, and the 64-PSK optical network in Fig. 5. Meanwhile, it is also found that the difference between the three carrier phase estimation methods becomes smaller for higher-level modulation formats.

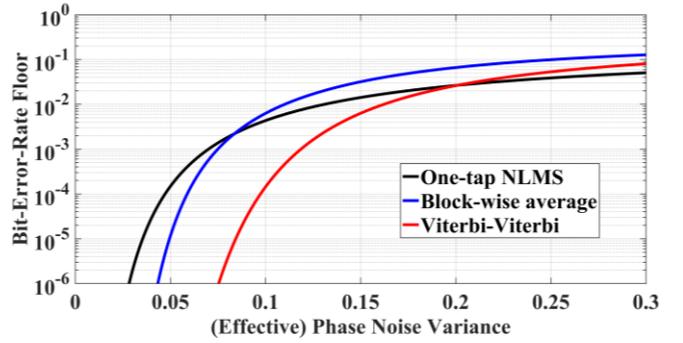

Fig. 2. BER floors versus PN variances in the three CPE approaches in 8-PSK optical fibre transmission network.

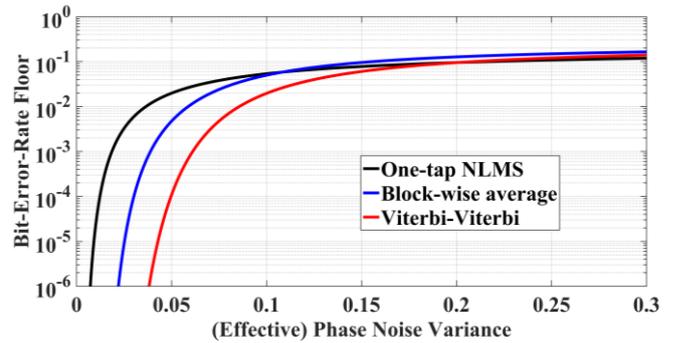

Fig. 3. BER floors versus PN variances in the three CPE approaches in 16-PSK optical fibre transmission network.

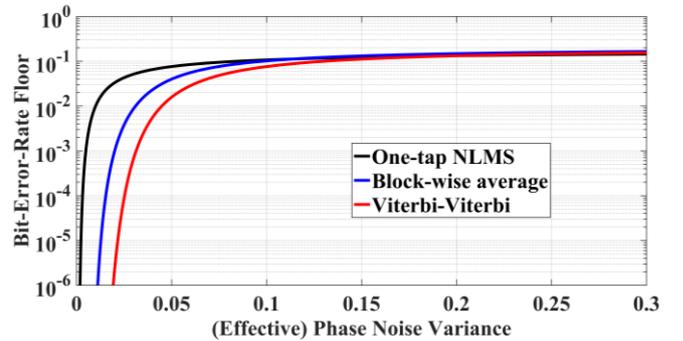

Fig. 4. BER floors versus PN variances in the three CPE approaches in 32-PSK optical fibre transmission network.

It is noted that the OT-NLMS algorithm can also be employed for the $n$-QAM coherent optical systems, while the block-wise average and the Viterbi-Viterbi methods cannot be easily used for the square $n$-QAM coherent systems except the circular constellation $n$-QAM systems.

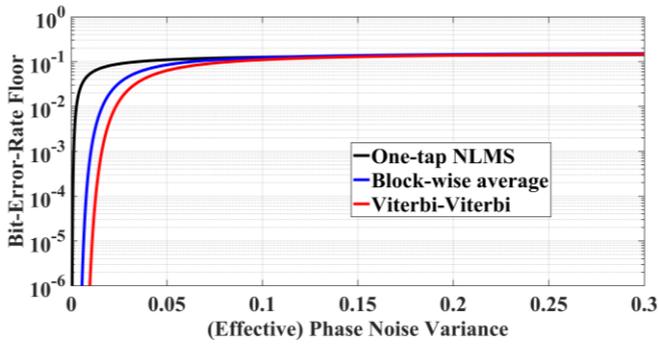

Fig. 5. BER floors versus PN variances in the three CPE approaches in 64-PSK optical fibre transmission network.

## V. CONCLUSION

The analysis of CPE in long-haul high-capacity optical communication networks, using the OT-NLMS, the BWA, and the VV approaches, are investigated, both considering laser PN and EEPN. The close-form description for estimating carrier phase in these carrier phase estimation approaches are analyzed, and the BER performance in these CPE approaches is also studied in detail using different modulation formats.

It is found that, VV CPE method outperforms the OT-NLMS and BWA approaches for a small PN variance (or effective PN variance), whereas three CPE algorithms will converge to a similar behavior for a large PN variance (or effective PN variance). Furthermore, the difference between the three CPE methods becomes smaller for higher-level modulation formats.

ACKNOWLEDGMENT

Financial supports under UK EPSRC program UNLOC EP/J017582/1, EU program GRIFFON 324391, EU program ICONE 608099, and Swedish Vetenskapsradet 0379801.